\documentclass[%
 reprint,
 amsmath,amssymb,
 aps,
]{revtex4-2}

\usepackage{graphicx}
\usepackage{dcolumn}
\usepackage{bm}
\usepackage{physics}
\usepackage{caption}
\usepackage{subcaption}
\usepackage{float}
\usepackage{hyperref}
\hypersetup{
    colorlinks=true,
    linkcolor=red,
    filecolor=magenta,
    urlcolor=cyan,
    citecolor=blue
}

\begin{document}

\preprint{APS/123-QED}

\title{Impact of independent reservoirs on the quantum Zeno and anti-Zeno effects}

\author{Irfan Javed}
\author{Mohsin Raza}%
\author{Adam Zaman Chaudhry}%
 \email{adam.zaman@lums.edu.pk}
 
\affiliation{%
 School of Science and Engineering, Lahore University of Management Sciences (LUMS), Opposite Sector U, D.H.A., Lahore 54792, Pakistan
}%

\date{\today}

\begin{abstract}
In this paper, we look into what happens to a quantum system under repeated measurements if it interacts with two independent reservoirs. In particular, we look at the behavior of a two-level system interacting with reservoirs consisting of harmonic oscillators. The interaction with one reservoir is weak with a dissipative-type coupling, while the interaction with the other reservoir is strong with a dephasing-type coupling. Using a polaron transformation, we show that the presence of the strongly coupled reservoir can actually reduce the decay rate of the quantum system due to the effect of the weakly-coupled reservoir. 
\end{abstract}

\maketitle


\section{Introduction}

The effect of quantum measurements on quantum systems is generally non-trivial. For example, repeated measurements on a quantum system can hinder the time evolution of the quantum system, an effect dubbed the quantum Zeno effect \cite{Sudarshan1977,FacchiPhysLettA2000,FacchiPRL2002,
FacchiJPA2008,WangPRA2008,ManiscalcoPRL2008,FacchiJPA2010,MilitelloPRA2011,RaimondPRA2012,SmerziPRL2012,
WangPRL2013,McCuskerPRL2013,StannigelPRL2014,ZhuPRL2014,SchafferNatCommun2014,SignolesNaturePhysics2014,
DebierrePRA2015,AlexanderPRA2015,QiuSciRep2015,HePRA2018,HanggiNJP2018,HePRA2019,MullerAnnPhys}. On the other hand, if the time interval between the measurements is not so short, the reverse effect, namely the quantum anti-Zeno effect, takes place whereby the evolution of the quantum system is actually accelerated\cite{KurizkiNature2000,RaizenPRL2001,BaronePRL2004,KoshinoPhysRep2005,BennettPRB2010,YamamotoPRA2010,ChaudhryPRA2014zeno,Chaudhryscirep2017b,HePRA2017,WuPRA2017,Chaudhryscirep2018,WuAnnals2018,ChaudhryEJPD2019a}. These effects have gained considerable interest both on the theoretical and experimental fronts not only due to huge importance in the foundations of quantum mechanics, but also their possible applications in quantum technologies. For example, the effect of repeated measurements can be used to infer properties of the environment of a quantum system, that is, noise sensing \cite{sakuldee2020,mullerPLA2020,sakuldeePRA2020}. Generally speaking, so far, the main focus of the studies performed on QZE and the QAZE have been on the population decay model \cite{KurizkiNature2000,RaizenPRL2001,BaronePRL2004,KoshinoPhysRep2005,ManiscalcoPRL2006,SegalPRA2007,ZhengPRL2008,BennettPRB2010,YamamotoPRA2010,AiPRA2010,ThilagamJMP2010,ThilagamJCP2013} and the dephasing model \cite{ChaudhryPRA2014zeno} (notable exceptions include Refs.~\cite{Chaudhryscirep2016,Chaudhryscirep2017a}), whereby the interplay between the measurements performed on a single two-level system and the effect of a single reservoir coupled to the two-level system is studied.

In this work, we look at the effect of repeated measurements on a quantum system which is interacting with two independent reservoirs. Such studies have gained interest lately due to the interplay between the effects of the independent reservoirs. For example, it has been shown that the effect of the two reservoirs are not merely additive in terms of the master equation describing the evolution of the central quantum system; in fact, there can be interference effects between the reservoirs \cite{ChanPRA2014,BylickaPRA2018}. Along these lines, in this paper, we look at a two-level system in the presence of two independent harmonic oscillator reservoirs. The interaction with one reservoir is weak, and is of the usual dissipative form inducing quantum decay from the excited state to the ground state. The second reservoir also consists of a collection of harmonic oscillators, and its interaction induces dephasing of the central two-level system. We then look at the effect of performing repeated measurements on the two-level system interacting with these two independent reservoirs. The effect of the strongly coupled reservoir can be handled exactly via a polaron transformation, while the weakly coupled reservoir is treated perturbatively. This treatment allows us to find analytical expressions for the decay rate of the quantum system, taking into account the repeated measurements. We find that the presence of the strongly coupled reservoir greatly changes the Zeno and anti-Zeno behavior. For example, the transition from the Zeno to the anti-Zeno regimes occurs for smaller measurement time intervals. Most interestingly, the presence of the strongly coupled reservoir, instead of enhancing the effect of the dissipative reservoir, actually reduces the effective decay rates.

\section{\label{sec. 1}Results}
We start with an extension of the paradigmatic spin-boson model \cite{LeggettRMP1987,Weissbook} that allows us to deal with two independent reservoirs. The system-environment Hamiltonian is 
\begin{equation}
\begin{aligned}[b]
H_{L}^{(0)} =\:&\frac{\epsilon}{2}\sigma_{z}+\frac{\Delta}{2}\sigma_{x}+\sum_{k}\omega_{k}b_{k}^{\dagger}b_{k}+\sum_{k}\alpha_{k}c_{k}^{\dagger}c_{k}\\
&+\sigma_{z}\sum_{k}\left(g_{k}^{*}b_{k}+g_{k}b_{k}^{\dagger}\right)+\sigma_{x}\sum_{k}\left(f_{k}^{*}c_{k}+f_{k}c_{k}^{\dagger}\right).
\end{aligned}
\label{eq. 1}
\end{equation}
Here, $\frac{\epsilon}{2}\sigma_{z}+\frac{\Delta}{2}\sigma_{x}$ is the system Hamiltonian, $\sum_{k}\omega_{k}b_{k}^{\dagger}b_{k}+\sum_{k}\alpha_{k}c_{k}^{\dagger}c_{k}$ is the total environment Hamiltonian, and $\sigma_{z}\sum_{k}\left(g_{k}^{*}b_{k}+g_{k}b_{k}^{\dagger}\right)+\sigma_{x}\sum_{k}\left(f_{k}^{*}c_{k}+f_{k}c_{k}^{\dagger}\right)$ gives the total interaction of the system with the two reservoirs. The energy-level splitting of the two-level system is $\epsilon$, while $\Delta$ is the tunneling amplitude, and $\omega_{k}$ and $\alpha_{k}$ are the frequencies of harmonic oscillators in the two harmonic oscillator reservoirs interacting with the system. $b_{k}/b_{k}^{\dagger}$ and $c_{k}/c_{k}^{\dagger}$ are the annihilation/creation operators of the first and second reservoirs, respectively, $\sigma_{x}$ and $\sigma_{z}$ are the standard Pauli operators, and we throughout we work in dimensionless units with $\hbar = 1$. The superscript $(0)$ simply means that we are dealing with the system-environment Hamiltonian that contains two reservoirs; later on, for comparison, we also consider only the single weakly-coupled reservoir, and the consequent system-environment Hamiltonian we denote as $H_L^{(1)}$. The subscript $L$ denotes the `lab' frame, that is, we have currently not performed any unitary transformation on the Hamiltonian. Henceforth, we use the following definitions: $H_{S1} = \frac{\epsilon}{2}\sigma_{z}$, $H_{S2} = \frac{\Delta}{2}\sigma_{x}$, $H_{B1} = \sum_{k}\omega_{k}b_{k}^{\dagger}b_{k}$, $H_{B2} = \sum_{k}\alpha_{k}c_{k}^{\dagger}c_{k}$, $V_{C1} = \sigma_{z}\sum_{k}\left(g_{k}^{*}b_{k}+g_{k}b_{k}^{\dagger}\right)$, and $V_{C2} = \sigma_{x}\sum_{k}\left(f_{k}^{*}c_{k}+f_{k}c_{k}^{\dagger}\right)$. We assume that $\abs{g_{k}}\gg\abs{f_{k}}$, that is, the coupling with the dephasing-type reservoir is much stronger than the dissipative coupling to the other reservoir. 

Our objective is to find the effective decay rate of the central two-level system in the presence of repeated projective measurements. To this end, we first endeavor to find the density matrix of the system as a function of time. We use a polaron transformation defined by $U_{P} = e^{\chi\sigma_{z}/2}$, where $\chi = \sum_{k}\left(\frac{2g_{k}}{\omega_{k}}b_{k}^{\dagger}-\frac{2g_{k}^{*}}{\omega_{k}}b_{k}\right)$~\cite{SilbeyJCP1984, VorrathPRL2005}. In the polaron frame, the Hamiltonian is
\begin{equation}
\begin{aligned}[b]
H^{(0)} =\:&\frac{\epsilon}{2}\sigma_{z}+\sum_{k}\omega_{k}b_{k}^{\dagger}b_{k}+\sum_{k}\alpha_{k}c_{k}^{\dagger}c_{k}\\
&+\left[\frac{\Delta}{2}+\sum_{k}\left(f_{k}^{*}c_{k}+f_{k}c_{k}^{\dagger}\right)\right]\left(\sigma_{+}e^{\chi}+\sigma_{-}e^{-\chi}\right).
\end{aligned}
\label{eq. 2}
\end{equation}
In the polaron frame then, the interaction of the system with the environment is weak, if we also assume the tunneling amplitude to be small, with the environment part of the coupling Hamiltonian modifying to $\left(\sigma_{+}e^{\chi}+\sigma_{-}e^{-\chi}\right)$. This weak coupling can then be treated perturbatively. The initial system-environment state can then be written as $\rho(0) = \rho_{S}(0)\otimes\rho_{B1}\otimes\rho_{B2}$, where $\rho_{S}(0) = \ket{\uparrow}\bra{\uparrow}$, $\rho_{B1}(0) = \frac{e^{-\beta H_{B1}}}{Z_{B1}}$ with $Z_{B1} = \mathrm{Tr}_{B1}\left[e^{-\beta H_{B1}}\right]$, and $\rho_{B2}(0) = \frac{e^{-\beta H_{B2}}}{Z_{B2}}$ with $Z_{B2} = \mathrm{Tr}_{B2}\left[e^{-\beta H_{B2}}\right]$. Using time-dependent perturbation theory to second-order in the polaron frame, we find that
\newpage
\begin{widetext}
\begin{equation}
\begin{aligned}[b]
\rho_{S}(\tau) = U_{S}(\tau)\bigg[&\rho_{S}(0)+i\sum_{\mu}\int_{0}^{\tau}dt_{1}\left[\rho_{S}(0), \widetilde{F}_{\mu}(t_{1})\right]\left\langle\widetilde{B}_{\mu}(t_{1})\right\rangle_{B1}\left\langle\widetilde{J}_{\mu}(t_{1})\right\rangle_{B2}\\
&+i\frac{\Delta}{2}\sum_{\mu}\int_{0}^{\tau}dt_{1}\left[\rho_{S}(0), \widetilde{F}_{\mu}(t_{1})\right]\left\langle\widetilde{B}_{\mu}(t_{1})\right\rangle_{B1}\\
&+\sum_{\mu\nu}\int_{0}^{\tau}dt_{1}\int_{0}^{t_{1}}dt_{2}\left(\left[\widetilde{F}_{\mu}(t_{1}), \rho_{S}(0)\widetilde{F}_{\nu}(t_{2})\right]C_{\mu\nu}(t_{1}, t_{2})K_{\mu\nu}(t_{1}, t_{2})+\mathrm{h.c.}\right)\\
&+\frac{\Delta}{2}\sum_{\mu\nu}\int_{0}^{\tau}dt_{1}\int_{0}^{t_{1}}dt_{2}\left(\left[\widetilde{F}_{\mu}(t_{1}), \rho_{S}(0)\widetilde{F}_{\nu}(t_{2})\right]C_{\mu\nu}(t_{1}, t_{2})\left\langle\widetilde{J}_{\nu}(t_{2})\right\rangle_{B2}+\mathrm{h.c.}\right)\\
&+\frac{\Delta}{2}\sum_{\mu\nu}\int_{0}^{\tau}dt_{1}\int_{0}^{t_{1}}dt_{2}\left(\left[\widetilde{F}_{\mu}(t_{1}), \rho_{S}(0)\widetilde{F}_{\nu}(t_{2})\right]C_{\mu\nu}(t_{1}, t_{2})\left\langle\widetilde{J}_{\mu}(t_{1})\right\rangle_{B2}+\mathrm{h.c.}\right)\\
&+\frac{\Delta^{2}}{4}\sum_{\mu\nu}\int_{0}^{\tau}dt_{1}\int_{0}^{t_{1}}dt_{2}\left(\left[\widetilde{F}_{\mu}(t_{1}), \rho_{S}(0)\widetilde{F}_{\nu}(t_{2})\right]C_{\mu\nu}(t_{1}, t_{2})+\mathrm{h.c.}\right)\bigg]U_{S}^{\dagger}(\tau).
\end{aligned}
\label{eq. 3}
\end{equation}
\end{widetext}
Here, $F_{1} = \sigma_{+}$, $F_{2} = \sigma_{-}$, $B_{1} = X$, $B_{2} = X^{\dagger}$, $J_{1} = J_{2} = \sum_{k}\left(f_{k}^{*}c_{k}+f_{k}c_{k}^{\dagger}\right)$, and h.c. denotes the Hermitian conjugate. As to their time-evolved counterparts, $\widetilde{F}_{\mu}(t) = U_{S}^{\dagger}(t)F_{\mu}U_{S}(t)$ with $U_{S}(t) = e^{-i H_{S1}t}$, $\widetilde{B}_{\mu}(t) = U_{B1}^{\dagger}(t)B_{\mu}U_{B1}(t)$ with $U_{B1}(t) = e^{-i H_{B1}t}$, and $\widetilde{J}_{\mu}(t) = U_{B2}^{\dagger}(t)J_{\mu}U_{B2}(t)$ with $U_{B2}(t) = e^{-i H_{B2}t}$. Finally, $\langle\ldots\rangle_{B}$ stands for $\mathrm{Tr}_{B}\left[\rho_{B}(\ldots)\right]$, and the environment correlation functions are defined as $C_{\mu\nu}(t_{1}, t_{2}) = \left\langle\widetilde{B}_{\mu}(t_{1})\widetilde{B}_{\nu}(t_{2})\right\rangle_{B1}$ and $K_{\mu\nu}(t_{1}, t_{2}) = \left\langle\widetilde{J}_{\mu}(t_{1})\widetilde{J}_{\nu}(t_{2})\right\rangle_{B2}$. At time $t = 0$, we prepare our system in the state $\ket{\uparrow}$, where $\sigma_{z}\ket{\uparrow} = \ket{\uparrow}$, and we subsequently perform a measurement after every interval of duration $\tau$ to check if the system state is still $\ket{\uparrow}$. The survival probability at time $t = \tau$ is then $s(\tau) = 1-\bra{\downarrow}\rho_{S}(\tau)\ket{\downarrow}$. Using our expression for $\rho_{S}(\tau)$ (Eq.~(\ref{eq. 3})), we obtain
\begin{widetext}
\begin{equation}
\begin{aligned}[b]
s(\tau) =\:&1-2\Re\left(\int_{0}^{\tau}dt_{1}\int_{0}^{t_{1}}dt_{2}C_{12}(t_{1}, t_{2})K_{12}(t_{1}, t_{2})e^{i\epsilon(t_{1}-t_{2})}\right)\\
&-\frac{\Delta^{2}}{2}\Re\left(\int_{0}^{\tau}dt_{1}\int_{0}^{t_{1}}dt_{2}C_{12}(t_{1}, t_{2})e^{i\epsilon(t_{1}-t_{2})}\right)\\
=\:&1-2\Re\left(\int_{0}^{\tau}dt_{1}\int_{0}^{t_{1}}dt_{2}C(t_{1}-t_{2})K(t_{1}-t_{2})e^{i\epsilon(t_{1}-t_{2})}\right)\\
&-\frac{\Delta^{2}}{2}\Re\left(\int_{0}^{\tau}dt_{1}\int_{0}^{t_{1}}dt_{2}C(t_{1}-t_{2})e^{i\epsilon(t_{1}-t_{2})}\right)\\
=\:&1-2\Re\left[\int_{0}^{\tau}dt\int_{0}^{t}dt'\left(C(t')K(t')e^{i\epsilon t'}+\frac{\Delta^{2}}{4}C(t')e^{i\epsilon t'}\right)\right].
\end{aligned}
\label{eq. 4}
\end{equation}
\end{widetext}
In the penultimate step, we use $C(t_{1}-t_{2}) = C_{12}(t_{1}, t_{2})$ and $K(t_{1}-t_{2}) = K_{12}(t_{1}, t_{2})$, and in the last step, we change variables via $t' = t_{1}-t_{2}$ and $t = t_{1}$. When we calculate the environment correlation functions, we find that $C(t) = e^{-\Phi_{R1}(t)}e^{-i\Phi_{I1}(t)}$ with $\Phi_{R1}(t) = 4\int_{0}^{\infty}d\omega J(\omega)\frac{1-\cos(\omega t)}{\omega^{2}}\coth(\frac{\beta\omega}{2})$ and $\Phi_{I1}(t) = 4\int_{0}^{\infty}d\omega J(\omega)\frac{\sin(\omega t)}{\omega^2}$ and that $K(t) = \Phi_{R2}(t)-i\Phi_{I2}(t)$ with $\Phi_{R2}(t) = \int_{0}^{\infty}d\alpha H(\alpha)\cos(\alpha t)\coth(\frac{\beta\alpha}{2})$ and $\Phi_{I2}(t) = \int_{0}^{\infty}d\alpha H(\alpha)\sin(\alpha t)$. Here, spectral densities have been introduced as $\sum_{k}\abs{g_k}^{2}(\ldots)\,\to\,\int_{0}^{\infty}d\omega J(\omega)(\ldots)$ and $\sum_{k}\abs{f_k}^{2}(\ldots)\,\to\,\int_{0}^{\infty}d\alpha H(\alpha)(\ldots)$. Since system-environment coupling in the polaron frame is weak, we could neglect the build-up of correlations between the system and the environment and write the survival probability at time $t = N\tau$ as $S(t = N\tau) = \left[s(\tau)\right]^{N} \equiv e^{-\Gamma^{(0)}(\tau)N\tau}$, thereby defining the effective decay rate $\Gamma^{(0)}(\tau)$~\cite{ChaudhryPRA2014zeno}. It follows that $\Gamma^{(0)}(\tau) = -\frac{1}{\tau}\ln s(\tau)$. Using the expression we derived for $s(\tau)$ (Eq.~(\ref{eq. 4})), we can work out the following expression for $\Gamma^{(0)}(\tau)$:
\begin{equation*}
\begin{aligned}[b]
\frac{2}{\tau}\int_{0}^{\tau}dt\int_{0}^{t}dt'\bigg(&e^{-\Phi_{R1}\left(t'\right)}\cos(\epsilon t'-\Phi_{I1}(t'))\Phi_{R2}(t')\\
&+e^{-\Phi_{R1}\left(t'\right)}\sin(\epsilon t'-\Phi_{I1}(t'))\Phi_{I2}(t')\\
&+\frac{\Delta^{2}}{4}e^{-\Phi_{R1}\left(t'\right)}\cos(\epsilon t'-\Phi_{I1}(t'))\bigg).
\end{aligned}
\end{equation*}

\begin{figure}[H]
\centering
\begin{subfigure}[H]{0.48\textwidth}
\centering
\includegraphics[width = \textwidth]{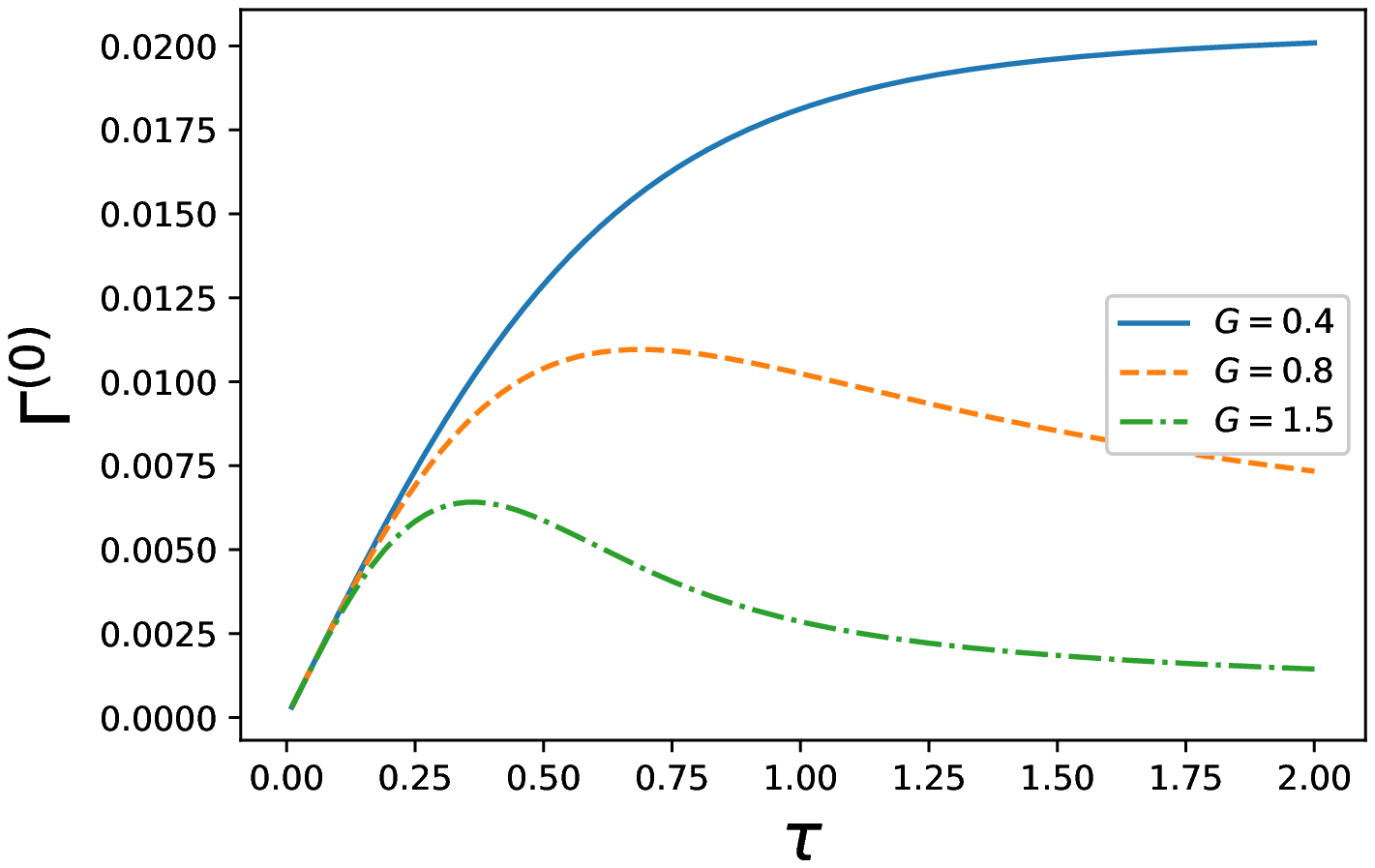}
\caption{}
\label{fig. 1a}
\end{subfigure}
\hfill
\begin{subfigure}[H]{0.48\textwidth}
\centering
\includegraphics[width = \textwidth]{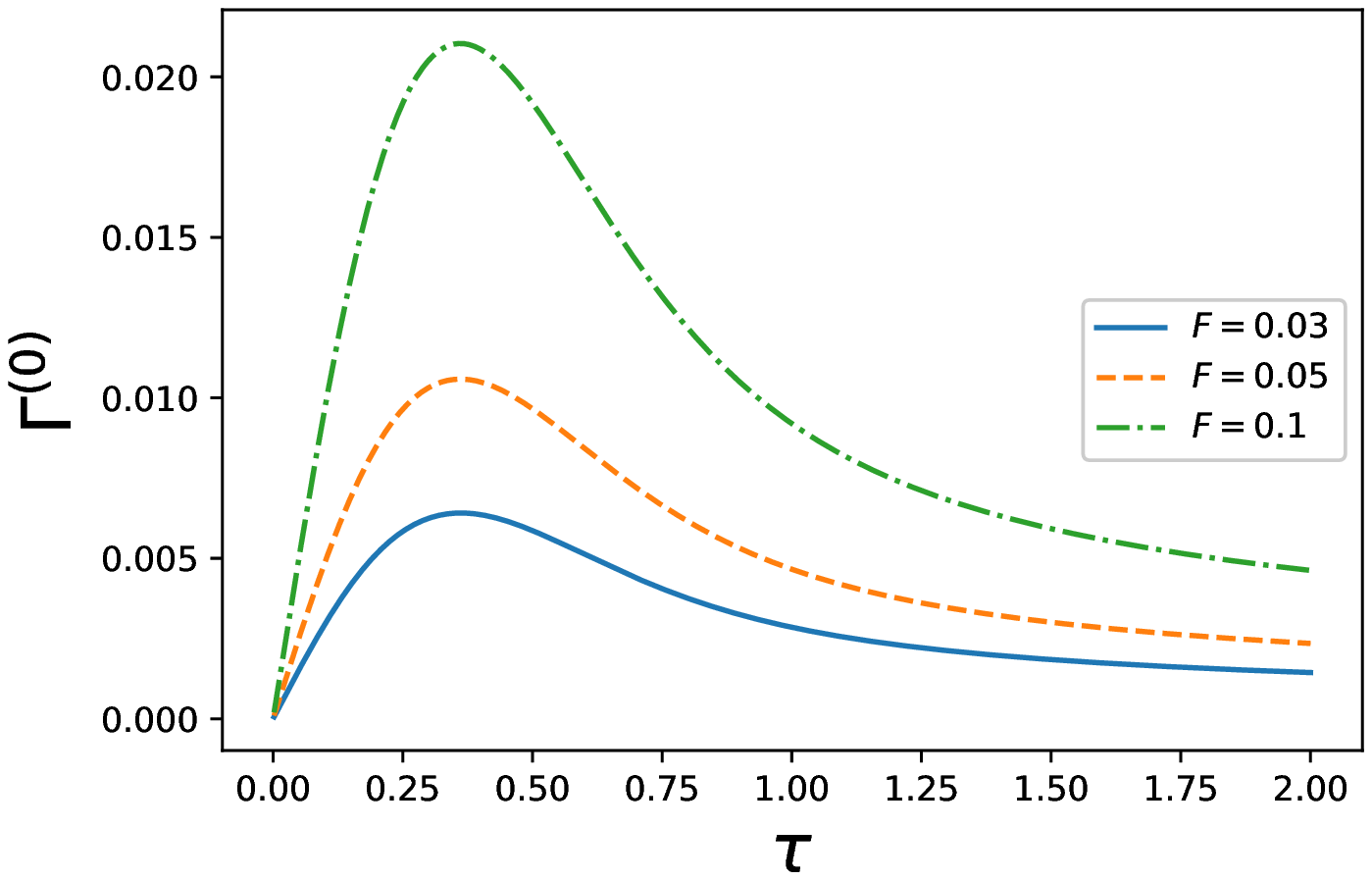}
\caption{}
\label{fig. 1b}
\end{subfigure}
\caption{\textbf{Variation of the effective decay rate with both strong and weak system-environment coupling strengths.} (a) Graph of $\Gamma^{(0)}$ (at zero temperature) against $\tau$ when $G = 0.4$ (solid, blue curve), $G = 0.8$ (dashed, orange curve), and $G = 1.5$ (dot-dashed, green curve). Here, we have used an Ohmic environment ($s = 1$ and $r = 1$) with $F = 0.03$, $\epsilon = 1$, $\omega_{c} = 1$, $\alpha_{c} = 1$, and $\Delta = 0.05$. The initial state is $\ket{\uparrow}$. (b) Graph of $\Gamma^{(0)}$ (at zero temperature) against $\tau$ when $F = 0.03$ (solid, blue curve), $F = 0.05$ (dashed, orange curve), and $F = 0.1$ (dot-dashed, green curve). Here, we have again used an Ohmic environment ($s = 1$ and $r = 1$) with $G = 1.5$, $\epsilon = 1$, $\omega_{c} = 1$, $\alpha_{c} = 1$, and $\Delta = 0.05$. The initial state is still $\ket{\uparrow}$. $\hbar$ is equal to $1$ throughout.}
\label{fig. 1}
\end{figure}

We now plot $\Gamma^{(0)}(\tau)$ against $\tau$. To do so, we model the spectral densities as $J(\omega) = G\omega^{s}\omega_{c}^{1-s}e^{-\omega/\omega_{c}}$ and $H(\alpha) = F\alpha^{r}\alpha_{c}^{1-r}e^{-\alpha/\alpha_{c}}$, where $G$ and $F$ are dimensionless parameters characterizing the system-environment coupling strengths, $\omega_{c}$ and $\alpha_{c}$ are cut-off frequencies, and $s$ and $r$ are Ohmicity parameters~\cite{BPbook}. The coupling strength $G$ corresponds to strongly coupled reservoir, while $F$ corresponds to the weakly coupled reservoir; therefore, we will consider $G$ to be significantly greater than $F$. To be particular, we work at zero temperature and look at the Ohmic case for each of the spectral densities ($s = 1$ and $r = 1$). Doing so gives $\Phi_{R1}(t) = 2G\ln(1+\omega_{c}^{2}t^{2})$, $\Phi_{I1}(t) = 4G\tan^{-1}(\omega_{c}t)$, $\Phi_{R2}(t) = F\frac{\alpha_{c}^2\left(1-\alpha_{c}^{2}t^{2}\right)}{\left(1+\alpha_{c}^{2}t^{2}\right)^{2}}$, and $\Phi_{I2}(t) = 2F\frac{\alpha_{c}^{3}t}{\left(1+\alpha_{c}^{2}t^{2}\right)^{2}}$, allowing us to write our expression for $\Gamma^{(0)}(\tau)$ as
\begin{equation}
\begin{aligned}[b]
&\Gamma^{(0)}(\tau) = \\
&\frac{2F}{\tau}\int_{0}^{\tau}dt\int_{0}^{t}dt'\frac{\alpha_{c}^{2}\left(1-\alpha_{c}^{2}t'^{2}\right)\cos(\epsilon t'-4G\tan^{-1}(\omega_{c}t'))}{\left(1+\alpha_{c}^{2}t'^{2}\right)^{2}\left(1+\omega_{c}^{2}t'^2\right)^{2G}}\\
&+\frac{4F}{\tau}\int_{0}^{\tau}dt\int_{0}^{t}dt'\frac{\alpha_{c}^{3}t'\sin(\epsilon t'-4G\tan^{-1}(\omega_{c}t'))}{\left(1+\alpha_{c}^{2}t'^{2}\right)^{2}\left(1+\omega_{c}^{2}t'^2\right)^{2G}}\\
&+\frac{2}{\tau}\int_{0}^{\tau}dt\int_{0}^{t}dt'\frac{\Delta^{2}}{4}\frac{\cos(\epsilon t'-4G\tan^{-1}(\omega_{c}t'))}{\left(1+\omega_{c}^{2}t'^2\right)^{2G}}.
\end{aligned}
\label{effdecayrate}
\end{equation}
While the effective decay rate has a relatively simple dependence on the coupling strength with weakly-coupled reservoir $F$, the dependence on the coupling strength with the strongly-coupled reservoir is very non-trivial. To investigate this dependence, the double integrals can be worked out numerically to obtain the effective decay rate as a function of the measurement interval $\tau$. Results are shown in Fig.~\ref{fig. 1} for different values of the system-environment coupling strengths, $G$ and $F$. Increasing the strong coupling strength clearly leads to a decrease in the decay rate (see Fig.~\ref{fig. 1a}). This behavior is in sharp contrast with what happens when the weak coupling strength is increased. Referring to Fig.~\ref{fig. 1b}, we note that increasing the weak coupling strength increases the decay rate. These behaviors are indicative of the fact that the strongly-coupled reservoir is actually inhibiting the decay rate induced due to the weakly coupled reservoir. It should also be noted that the behavior of the decay rate as a function of the measurement interval, $\tau$, allows us to identify the Zeno and anti-Zeno regimes. One way is to simply state that we are in the Zeno regime if the decay rate decreases when $\tau$ decreases and that we are in the anti-Zeno regime if the decay rate increases when $\tau$ decreases~\cite{ChaudhryPRA2014zeno, Chaudhryscirep2016, WuAnnals2018}. As is evident in Fig.~\ref{fig. 1}, whereas increasing the weak coupling strength does not change the qualitative behavior of the Zeno/anti-Zeno transition, increasing the strong coupling strength does have an effect, namely that it causes the transition to occur at smaller values of $\tau$. 
\vfill

\begin{figure}[b]
\centering
\begin{subfigure}[H]{0.48\textwidth}
\centering
\includegraphics[width = \textwidth]{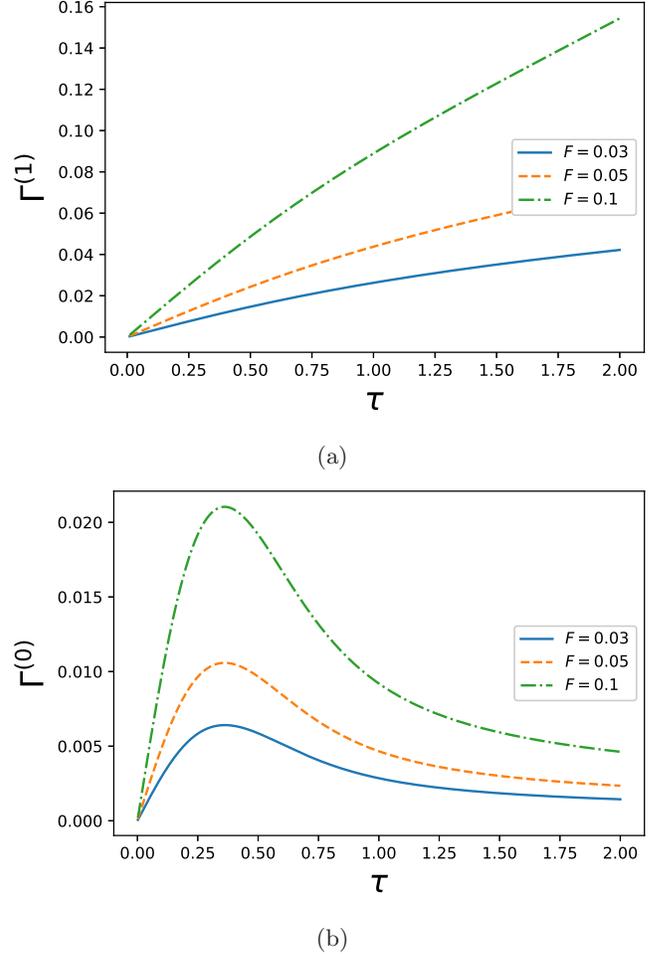}
\caption{}
\label{fig. 2a}
\end{subfigure}
\hfill
\begin{subfigure}[H]{0.48\textwidth}
\centering
\includegraphics[width = \textwidth]{set_2.eps}
\caption{}
\label{fig. 2b}
\end{subfigure}
\caption{\textbf{Variation of the effective decay rate with only weak system-environment coupling strength.} (a) Graph of $\Gamma^{(1)}$ (at zero temperature) against $\tau$ when $F = 0.03$ (solid, blue curve), $F = 0.05$ (dashed, orange curve), and $F = 0.1$ (dot-dashed, green curve). Here, we have used an Ohmic environment ($r = 1$) with $\epsilon = 1$, $\alpha_{c} = 1$, and $\Delta = 0.05$. The initial state is $\ket{\uparrow}$. (b) Graph of $\Gamma^{(0)}$ (at zero temperature) against $\tau$ when $F = 0.03$ (solid, blue curve), $F = 0.05$ (dashed, orange curve), and $F = 0.1$ (dot-dashed, green curve). Here, we have again used an Ohmic environment ($s = 1$ and $r = 1$) with $G = 1.5$, $\epsilon = 1$, $\omega_{c} = 1$, $\alpha_{c} = 1$, and $\Delta = 0.05$. The initial state is still $\ket{\uparrow}$. $\hbar$ is equal to $1$ throughout.}
\label{fig. 2}
\end{figure}
Having formulated the effective decay rate in the presence of both the strongly-coupled reservoir and the weakly-coupled reservoir, we now compute the effective decay in the presence of only the weakly-coupled dissipative reservoir for a better comparison. The system-environment Hamiltonian is now
\begin{equation}
H_{L}^{(1)} = \frac{\epsilon}{2}\sigma_{z}+\frac{\Delta}{2}\sigma_{x}+\sum_{k}\alpha_{k}c_{k}^{\dagger}c_{k}+\sigma_{x}\sum_{k}(f_{k}^{*}c_{k}+f_{k}c_{k}^{\dagger}),
\label{eq. 5}
\end{equation}
which differs from $H_{L}^{(0)}$ in Eq.~\eqref{eq. 1} in that it does not include $H_{B1}$ and $V_{C1}$. The superscript $(1)$ denotes that this is a variant of our original Hamiltonian, and all other symbols have the meanings ascribed to them previously. We rewrite $H_{L}^{(1)}$ as 
$$ H_L^{(1)} = \frac{\epsilon}{2}\sigma_{z}+\sum_{k}\alpha_{k}c_{k}^{\dagger}c_{k}+\left[\frac{\Delta}{2}+\sum_{k}\left(f_{k}^{*}c_{k}+f_{k}c_{k}^{\dagger}\right)\right]\sigma_{x}, $$
 and treat $\left[\frac{\Delta}{2}+\sum_{k}\left(f_{k}^{*}c_{k}+f_{k}c_{k}^{\dagger}\right)\right]\sigma_{x}$ as a weak perturbation in order to make a fair comparison with our previous case in which the strongly-coupled reservoir was also present. We now do not need to perform any unitary transformation since the system-environment interaction is weak in the lab frame itself. Writing the initial system-environment state as $\rho_{L}(0) = \rho_{S}(0)\otimes\rho_{B2}(0)$, where $\rho_{S}(0) = \ket{\uparrow}\bra{\uparrow}$ and $\rho_{B2}(0) = \frac{e^{-\beta H_{B2}}}{Z_{B2}}$ with $Z_{B2} = \mathrm{Tr}_{B2}\left[e^{-\beta H_{B2}}\right]$, and using time-dependent perturbation theory, we now find that
\begin{widetext}
\begin{equation}
\begin{aligned}[b]
\rho_{S}(\tau) = U_{S}(\tau)\bigg[&\rho_{S}(0)+i\frac{\Delta}{2}\sum_{\mu}\int_{0}^{\tau}dt_{1}\left[\rho_{S}(0), \widetilde{F}_{\mu}(t_{1})\right]\\
&+\sum_{\mu\nu}\int_{0}^{\tau}dt_{1}\int_{0}^{t_{1}}dt_{2}\left(\left[\widetilde{F}_{\mu}(t_{1}), \rho_{S}(0)\widetilde{F}_{\nu}(t_{2})\right]K_{\mu\nu}(t_{1}, t_{2})+\mathrm{h.c.}\right)\\
&+\frac{\Delta^{2}}{4}\sum_{\mu\nu}\int_{0}^{\tau}dt_{1}\int_{0}^{t_{1}}dt_{2}\left(\left[\widetilde{F}_{\mu}(t_{1}), \rho_{S}(0)\widetilde{F}_{\nu}(t_{2})\right]+\mathrm{h.c.}\right)\bigg]U_{S}^{\dagger}(\tau).
\end{aligned}
\label{eq. 6}
\end{equation}
\end{widetext}
Here, $F_{1} = \sigma_{+}$, $F_{2} = \sigma_{-}$, $J_{1} = \sum_{k}\left(f_{k}^{*}c_{k}+f_{k}c_{k}^{\dagger}\right)$, $J_{2} = \sum_{k}\left(f_{k}^{*}c_{k}+f_{k}c_{k}^{\dagger}\right)$, $\widetilde{F}_{\mu}(t) = U_{S}^{\dagger}(t)F_{\mu}U_{S}(t)$ with $U_{S}(t) = e^{-i H_{S1}t}$, $\widetilde{J}_{\mu}(t) = U_{B2}^{\dagger}(t)J_{\mu}U_{B2}(t)$ with $U_{B2}(t) = e^{-i H_{B2}t}$, $\langle\ldots\rangle_{B} = \mathrm{Tr}_{B}\left[\rho_{B}(\ldots)\right]$, and the environment correlation functions are defined as $K_{\mu\nu}(t_{1}, t_{2}) = \left\langle\widetilde{J}_{\mu}(t_{1})\widetilde{J}_{\nu}(t_{2})\right\rangle_{B2}$. Here we have used the fact that $\langle \widetilde{J}_\mu(t)\rangle_{B} = 0$ for our case. Once again, we prepare our system in the initial state $\ket{\uparrow}$ and perform a projective measurement after every a time interval of $\tau$ to check if the system state is still in the state $\ket{\uparrow}$. The survival probability at time $t = \tau$ is then $s(\tau) = 1-\bra{\downarrow}\rho_{S}(\tau)\ket{\downarrow}$. With the expression we found for $\rho_{S}(\tau)$ in Eq.~(\ref{eq. 6}), we obtain now
\begin{widetext}
\begin{equation}
\begin{aligned}[b]
s(\tau) =\:&1-2\Re\left(\int_{0}^{\tau}dt_{1}\int_{0}^{t_{1}}dt_{2}K_{12}(t_{1}, t_{2})e^{i\epsilon(t_{1}-t_{2})}\right)-\frac{\Delta^{2}}{2}\Re\left(\int_{0}^{\tau}dt_{1}\int_{0}^{t_{1}}dt_{2}e^{i\epsilon(t_{1}-t_{2})}\right)\\
=\:&1-2\Re\left(\int_{0}^{\tau}dt_{1}\int_{0}^{t_{1}}dt_{2}K(t_{1}-t_{2})e^{i\epsilon(t_{1}-t_{2})}\right)-\frac{\Delta^{2}}{2}\Re\left(\int_{0}^{\tau}dt_{1}\int_{0}^{t_{1}}dt_{2}e^{i\epsilon(t_{1}-t_{2})}\right)\\
=\:&1-2\Re\left[\int_{0}^{\tau}dt\int_{0}^{t}dt'\left(K(t')e^{i\epsilon t'}+\frac{\Delta^{2}}{4}e^{i\epsilon t'}\right)\right].
\end{aligned}
\label{eq. 7}
\end{equation}
\end{widetext}
In the penultimate step, we use $K(t_{1}-t_{2}) = K_{12}(t_{1}, t_{2})$, and in the last step, we change variables via $t' = t_{1}-t_{2}$ and $t = t_{1}$. Calculating the correlation function $K(t)$, we find that $K(t) = \Phi_{R2}(t)-i\Phi_{I2}(t)$ with $\Phi_{R2}(t) = \int_{0}^{\infty}d\alpha H(\alpha)\cos(\alpha t)\coth(\frac{\beta\alpha}{2})$ and $\Phi_{I2}(t) = \int_{0}^{\infty}d\alpha H(\alpha)\sin(\alpha t)$, where, as before, the spectral density has been introduced as $\sum_{k}\abs{f_k}^{2}(\ldots)\,\to\,\int_{0}^{\infty}d\alpha H(\alpha)(\ldots)$. Since system-environment coupling in the lab frame is weak, we can ignore the correlations building between the system and the environment and follow the reasoning used before to show that the effective decay rate, $\Gamma^{(1)}(\tau)$, is
\begin{equation*}
\begin{aligned}[b]
&\Gamma^{(1)}(\tau) = \\
&\frac{2}{\tau}\int_{0}^{\tau}dt\int_{0}^{t}dt'\bigg(\cos(\epsilon t')\Phi_{R2}(t')+\sin(\epsilon t')\Phi_{I2}(t')\\
&+\frac{\Delta^{2}}{4}\cos(\epsilon t')\bigg).
\end{aligned}
\end{equation*}
To plot $\Gamma^{(1)}(\tau)$ against $\tau$, we model the spectral density of the weak coupling bath the same way as before and look at the Ohmic case. Working at zero temperature then allows us to write our expression for $\Gamma^{(1)}(\tau)$ as
\begin{equation*}
\begin{aligned}[b]
&\Gamma^{(1)}(\tau) = \\
&\frac{2F}{\tau}\int_{0}^{\tau}dt\int_{0}^{t}dt'\frac{\alpha_{c}^{2}\left(1-\alpha_{c}^{2}t^{2}\right)\cos(\epsilon t')}{\left(1+\alpha_{c}^{2}t^{2}\right)^{2}}\\
&+\frac{4F}{\tau}\int_{0}^{\tau}dt\int_{0}^{t}dt'\frac{\alpha_{c}^{3}t\sin(\epsilon t')}{\left(1+\alpha_{c}^{2}t^{2}\right)^{2}}\\
&+\frac{2}{\tau}\int_{0}^{\tau}dt\int_{0}^{t}dt'\frac{\Delta^{2}}{4}\cos(\epsilon t').
\end{aligned}
\end{equation*}
Notice that again there is a relatively simple dependence on the coupling stregth $F$. To investigate the effective decay rate as a function of the measurement interval now, we evaluate the double integral numerically to plot $\Gamma^{(1)}(\tau)$ as a function of the measurement interval. Results are shown in Fig.~\ref{fig. 2}, where Fig.~\ref{fig. 2a} shows the decay rate without the strongly-coupled reservoir, while Fig.~\ref{fig. 2b} is with the strongly-coupled reservoir. It is evident that, as expected, increasing the weak coupling strength increases the effective decay rate. More importantly, comparing the values of the effective decay rates obtained in Figs.~\ref{fig. 2a} and \ref{fig. 2b}, it is clear that the presence of the strong-coupled reservoir, instead of enhancing the decay rate, inhibits the effect of the weakly-coupled dissipative reservoir.

\section{\label{sec. 3}Conclusion}

In conclusion, our study shows that the effect of two independent reservoirs on a quantum system can be very rich and complicated. In particular, we have focused on calculating the effective decay rate for a single two-level system coupled to two independent reservoirs consisting on a collection of harmonic oscillations, one with strong coupling strength, and the other with weak coupling strength. We have shown that the presence of the strongly-coupled reservoirs can reduce the effect of the weakly-coupled reservoir on the two-level system. This means that the effect of the two reservoirs does not simply add. Rather, since the two reservoirs interact with a common two-level system, there is an indirect interaction between the two reservoirs. This interplay is responsible for the presence of the strongly-coupled reservoir reducing the effective decay rate of the two-level system. Our results should be useful in quantum control in scenarios where the aim is to control a quantum system interacting with multiple reservoirs. 

\begin{acknowledgments}
We would like to extend sincere gratitude to our colleague Hudaiba Soomro for her unstinting support throughout the project.
\end{acknowledgments}


%

\end{document}